\documentclass[conference]{IEEEtran}
\usepackage{cite}
\usepackage[cmex10]{amsmath}
\usepackage{array}
\usepackage{amsfonts}
\usepackage{mathrsfs}
\usepackage{arydshln}
\usepackage{slashbox}
\usepackage{graphicx}
\usepackage{float}
\usepackage{subfigure}
\usepackage{amssymb}
\usepackage{amsmath}
\usepackage[colorlinks,linkcolor=black,urlcolor=black,anchorcolor=black,citecolor=black,hyperfootnotes=true]{hyperref}
\usepackage{multirow}
\usepackage{multicol}
\usepackage{cite}
\usepackage[subfigure]{graphfig}
\usepackage{xcolor}
\usepackage{setspace}
\newcommand{\mv}[1]{\mbox{\boldmath{$ #1 $}}}
\usepackage[linesnumbered,ruled]{algorithm2e}

\newtheorem{lemma}{\underline{Lemma}}
\newtheorem{proposition}{\underline{Proposition}}

\newcommand{\qed}{\nobreak \ifvmode \relax \else
      \ifdim\lastskip<1.5em \hskip-\lastskip
      \hskip1.5em plus0em minus0.5em \fi \nobreak
      \vrule height0.75em width0.5em depth0.25em\fi}

\begin{document}
\title{{Trajectory Design for Cellular-Connected UAV Under Outage Duration Constraint}
\author{\IEEEauthorblockN{Shuowen~Zhang and Rui~Zhang}
	\IEEEauthorblockA{ECE Department, National University of Singapore. Email: \{elezhsh,elezhang\}@nus.edu.sg}}}
\maketitle

\begin{abstract}
In this paper, we study the trajectory design for a cellular-connected unmanned aerial vehicle (UAV) with given initial and final locations, while communicating with the ground base stations (GBSs) along its flight. We consider delay-limited communications between the UAV and its associated GBSs, where a given signal-to-noise ratio (SNR) target needs to be satisfied at the receiver. However, in practice, due to various factors such as quality-of-service (QoS) requirement, GBSs' availability and UAV mobility constraints, the SNR target may not be met at certain time periods during the flight, each termed as an \emph{outage duration}. In this paper, we aim to optimize the UAV trajectory to minimize its mission completion time, subject to a constraint on the maximum tolerable outage duration in its flight. To tackle this non-convex problem, we first transform it into a more tractable form and thereby reveal some useful properties of the optimal trajectory solution. Based on these properties, we then further simplify the problem and propose efficient algorithms to check the feasibility of the problem as well as to obtain its optimal and high-quality suboptimal solutions, by leveraging graph theory and convex optimization techniques. Numerical results show that our proposed trajectory designs outperform the conventional method based on dynamic programming, in terms of both performance and complexity.
\end{abstract}

\vspace{-1mm}
\section{Introduction}
\vspace{-1mm}
Unmanned aerial vehicles (UAVs) are promising solutions for various applications such as cargo delivery, aerial inspection, video streaming, emergency response, etc., thanks to their high mobility and flexible deployment \cite{cellularUAV_arXiv}. With the dramatically increasing demand for UAVs, it is of paramount importance to ensure that all UAVs can operate safely and efficiently, which calls for high-quality communications between UAVs and their ground pilots/users. A new and cost-effective approach to achieve this goal is \emph{cellular-enabled UAV communication}, where the ground base stations (GBSs) in the \hbox{cellular network} are leveraged to communicate with UAVs by treating them as new \emph{aerial users} \cite{cellularUAV_arXiv,LTEsky,LinSky,3GPPUAV,multibeam,NOMA,Disconnectivity}. Compared to the traditional point-to-point UAV-ground communications via Wi-Fi which are restricted to the visual line-of-sight (VLoS) range, cellular-enabled UAV communication supports beyond VLoS (BVLoS) UAV operation by exploiting the high-speed backhaul links in the cellular network \cite{cellularUAV_arXiv}.

\begin{figure}[t]
	\centering
	\includegraphics[width=6cm]{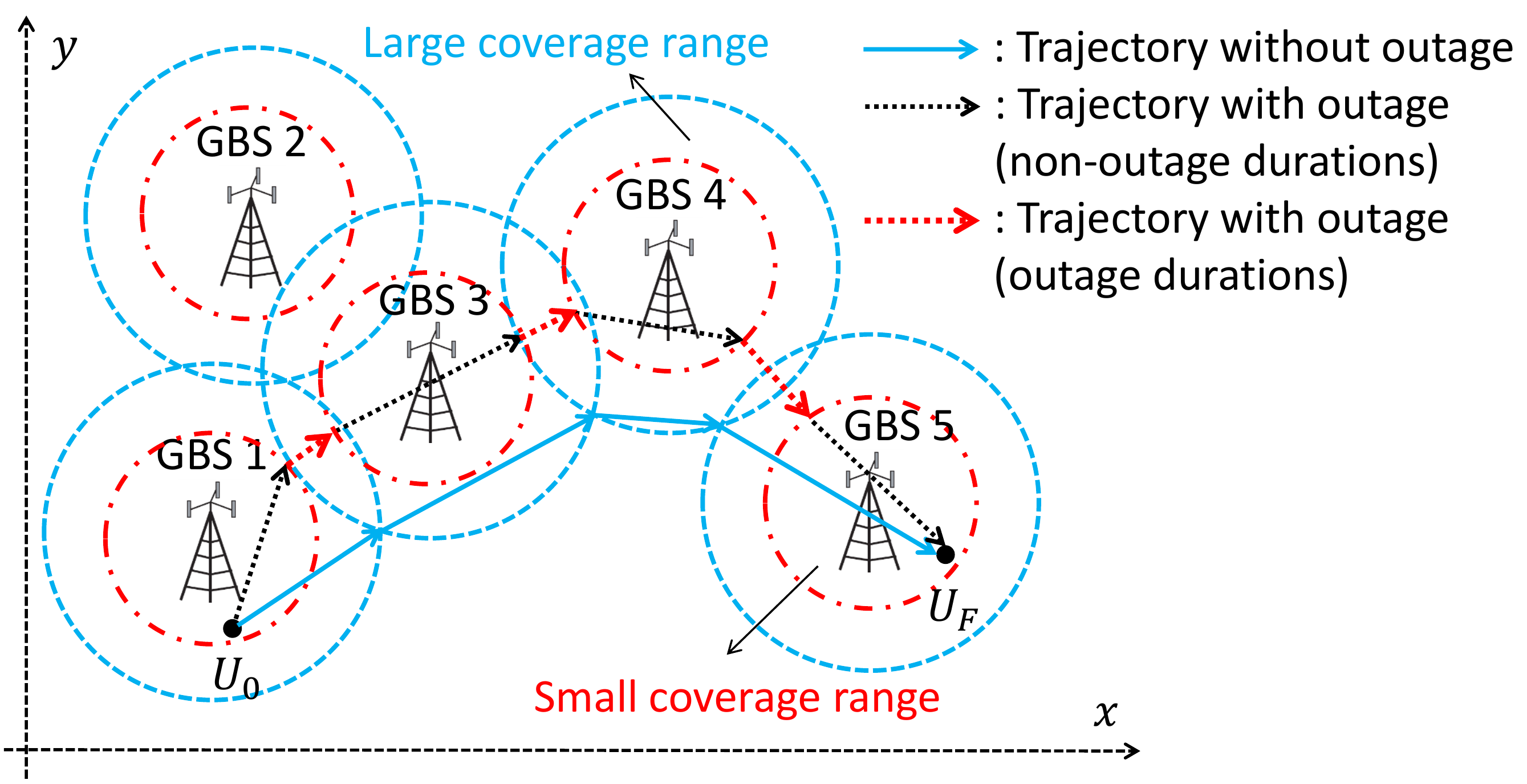}
	\vspace{-3mm}
	\caption{Illustration of trajectories without versus with communication outage.}\label{outage}
	\vspace{-8mm}
\end{figure}

Compared to traditional terrestrial users, UAVs have different channel characteristics with their serving GBSs, which give rise to both challenges and opportunities in the design of cellular-enabled UAV communications. Specifically, the communication channels between UAVs and GBSs are generally dominated by the line-of-sight (LoS) paths, which lead to a pronounced macro-diversity gain in associating UAVs with more GBSs with strong LoS channels as compared to the terrestrial users. However, on the other hand, LoS channels also incur severe interference with the non-associated GBSs. To resolve this issue, effective air-ground interference management techniques need to be devised \cite{multibeam,NOMA}. Among others, exploiting the UAV's high mobility in three-dimensional (3D) space to design its trajectory for enhancing the communication performances with GBSs is a promising new approach, which has been recently investigated in \cite{cellularUAV_arXiv,Disconnectivity}. Specifically, our prior work \cite{cellularUAV_arXiv} studied the trajectory design for a cellular-connected UAV in the mission of flying from an initial location to a final location, while communicating with its associated GBSs along the trajectory under a prescribed quality-of-service (QoS) requirement in terms of a minimum signal-to-noise ratio (SNR) at the receiver, which needs to be satisfied at all time. Under this stringent constraint, the UAV trajectory was optimized in \cite{cellularUAV_arXiv} to minimize the mission completion time.

However, in practice, due to various factors such as QoS requirement, GBSs' availability and UAV mobility constraints, a constant SNR target may not be met at certain time periods during the UAV's flight, each of which is termed as an \emph{outage duration}. For example, in Fig. \ref{outage}, we compare the two cases where there is no communication outage along the UAV trajectory (the setup considered in \cite{cellularUAV_arXiv}) versus that there are outage durations due to e.g., the increased SNR target (or equivalently, the reduced coverage range of each GBS).

In this paper, we extend our work in \cite{cellularUAV_arXiv} to the more challenging scenario where outage durations are inevitable in the UAV's flight. Specifically, we optimize the UAV trajectory to minimize its mission completion time from given initial to final locations, subject to a new constraint on the maximum tolerable outage duration over the flight. Note that for the special case of zero outage duration constraint, this problem reduces to that considered in \cite{cellularUAV_arXiv}. However, for the general case where finite outage duration is considered, the problem becomes more challenging to solve as compared to that in \cite{cellularUAV_arXiv}. As a result, the solution in \cite{cellularUAV_arXiv} cannot be applied to solve our considered problem in this paper under the general setup. To tackle the new problem,
we first transform it to an equivalent problem in a more tractable form, based on which useful structural properties of the optimal trajectory solution are derived. By leveraging these properties, we further simplify the problem and propose efficient algorithms to check its feasibility and obtain both its optimal as well as high-quality suboptimal solutions, by applying graph theory and convex optimization techniques. It is worth noting that the problem considered in this paper has also been studied in \cite{Disconnectivity}, where a dynamic programming (DP) based method was proposed to find an approximate solution. It is shown in this paper that our proposed trajectory designs outperform the DP-based solution in terms of both performance and complexity, thanks to the joint exploitation of graph theory and convex optimization in our proposed designs.

\vspace{-1mm}
\section{System Model}\label{sec_system}
\vspace{-1mm}
Consider a cellular-connected UAV and $M\geq 1$ GBSs that may potentially be associated with the UAV during its flight mission. We assume that the UAV flies at a constant altitude of $H$ meters (m), and all the $M$ GBSs have the same height of $H_{\mathrm{G}}$ m, with $H_{\mathrm{G}}\ll H$. The mission of the UAV is to fly from an initial location $U_0$ to a final location $U_F$, while communicating with the cellular network. By considering a 3D Cartesian coordinate system, we denote $(x_0,y_0,H)$ and $(x_F,y_F,H)$ as the coordinates of $U_0$ and $U_F$, respectively; $(a_m,b_m,H_{\mathrm{G}})$ as the coordinate of each $m$th GBS; and $(x(t),y(t),H),\ 0\leq t\leq T$ as the time-varying coordinate of the UAV, with $T$ denoting the mission completion time. For ease of exposition, we further define ${\mv{u}}_0=[x_0,y_0]^T$, ${\mv{u}}_F=[x_F,y_F]^T$, ${\mv{g}}_m=[a_m,b_m]^T$, and ${\mv{u}}(t)=[x(t),y(t)]^T$ to represent the above coordinates projected on the horizontal plane, respectively, where ${\mv{u}}(0)={\mv{u}}_0$ and ${\mv{u}}(T)={\mv{u}}_F$.

We assume that the channel between the UAV and each GBS is dominated by the LoS link. We also consider that the UAV is equipped with one single antenna, while each GBS is equipped with multiple antennas that have a fixed directional gain towards the UAV and hence can be equivalently treated as a single antenna for simplicity. Note that at each time instant $t$, the distance between the $m$th GBS and \hbox{the UAV is given by}
\vspace{-1mm}\begin{equation}
d_m(t)=\sqrt{(H-H_{\mathrm{G}})^2+\|{\mv{u}}(t)-{\mv{g}}_m\|^2},\quad m\in \mathcal{M},
\vspace{-1mm}\end{equation}
where $\|\cdot\|$ denotes the Euclidean norm, and $\mathcal{M}=\{1,...,M\}$ denotes the GBS index set. Therefore, the channel coefficient between the $m$th GBS and the UAV at time $t$ is expressed as
\vspace{-1mm}\begin{equation}\label{channel}
h_m(t)=\sqrt{\beta_0/d_m^2(t)}e^{-j\frac{2\pi}{\lambda}d_m(t)},\quad m\in\mathcal{M},
\vspace{-1mm}\end{equation}
where $\beta_0$ denotes the channel power gain at the reference distance of $d_0=1$ m, and $\lambda$ denotes the wavelength in m. We assume that the UAV is associated with one GBS indexed by $I(t)\in\mathcal{M}$ at each time instant $t$ during its mission. For convenience, we consider the scenario of downlink communication from the GBS to the UAV, while the results of this paper are also applicable to the uplink communication. The received signal at the UAV at one particular symbol interval can be expressed as
\vspace{-1mm}\begin{equation}
y=\sqrt{P}h_{I(t)}(t)s+z,\quad 0\leq t\leq T,
\vspace{-1mm}\end{equation}
where $P$ denotes the transmission power at GBS $I(t)$; $s$ denotes the information symbol sent by GBS $I(t)$, which is assumed to be a random variable with zero mean and unit variance; and $z\sim \mathcal{CN}(0,\sigma^2)$ denotes the circularly symmetric complex Gaussian (CSCG) noise with zero mean and variance $\sigma^2$. For simplicity, we assume that a dedicated time-frequency channel is assigned to the UAV communication, and hence there is no interference from other non-associated GBSs. According to (\ref{channel}), the GBS that is \emph{closest to} the UAV at each time instant $t$ yields the maximum channel power gain with the UAV, thus should be associated with the UAV for communication, i.e., $I(t)=\arg\underset{m\in\mathcal{M}}{\min}\|{\mv{u}}(t)-{\mv{g}}_m\|$. Consequently, the SNR at the UAV receiver at each time instant $t$ is given by
\vspace{-1mm}\begin{equation}\label{SINR}
\rho(t)=\frac{\rho_0}{(H-H_\mathrm{G})^2+\underset{m\in\mathcal{M}}{\min}\ \|{\mv{u}(t)-{\mv{g}}_m\|^2}},\ 0\leq t\leq T,
\vspace{-2mm}\end{equation}
where $\rho_0=\frac{P\beta_0}{\sigma^2}$ denotes the reference SNR at  \hbox{$d_0=1$ m.}

We consider delay-limited communications between the GBS and UAV for e.g., exchanging time-critical command and control (C$\&$C) messages, real-time video streaming, and so on. In practice, this type of communications generally requires a minimum SNR target to be satisfied at the receiver to meet the prescribed QoS requirements, namely $\rho(t)\geq \bar{\rho}$, where $\bar{\rho}$ denotes the SNR target. It can be shown from (\ref{SINR}) that this requirement is equivalent to the following constraint on the horizontal distance between the UAV and its associated (closest) GBS at each time instant $t$:
\vspace{-1mm}\begin{equation}\label{d}
\underset{m\in\mathcal{M}}{\min}\ \|{\mv{u}(t)-{\mv{g}}_m\|}\leq \bar{d},
\vspace{-1mm}\end{equation}
with $\bar{d}\!\overset{\Delta}{=}\!\sqrt{\frac{\rho_0}{\bar{\rho}}\!-\!(H\!-\!H_{\mathrm{G}})^2}$. Clearly, it is desirable to design the UAV trajectory $\{{\mv{u}}(t),0\leq t\leq T\}$ such that (\ref{d}) is satisfied for all time instants $t\!\in\! [0,T]$ throughout its mission, as pursued in our previous work \cite{cellularUAV_arXiv}. However, as discussed in Section I, this may not be always feasible in practice, since the existence of such trajectory depends on various factors such as the required communication range $\bar{d}$, the number of GBSs and their locations, etc. For any given UAV trajectory, an \emph{outage} event for UAV communication occurs at time $t$ \hbox{if (\ref{d}) is not satisfied.}

Motivated by the above practical issue in UAV trajectory design, we consider the \emph{maximum outage duration} in this paper. Specifically, for each time instant $t$ during the UAV mission, denote $t_{\mathrm{N}}(t)$, $t_{\mathrm{N}}(t)\leq t$, as the latest time instant at which there is no outage, i.e.,
\vspace{-1mm}
\begin{equation}\label{t_c}
t_{\mathrm{N}}(t)\!=\!\max\!\big\{\hat{t}\!\in\![0,t]\!:\!\underset{m\in \mathcal{M}}{\min}\ \|{\mv{u}}(\hat{t})\!-\!{\mv{g}}_m\|\!\leq\! \bar{d}\big\},0\!\leq\! t\!\leq\! T.
\vspace{-1mm}
\end{equation}
Note that $t_{\mathrm{N}}(t)=t$ if there is no outage at time $t$, while $t_{\mathrm{N}}(t)<t$ represents that outage occurs from $t_{\mathrm{N}}(t)$ to $t$ for a finite duration of $t-t_{\mathrm{N}}(t)$. The maximum outage duration over the UAV mission is thus given by 
\vspace{-2mm}\begin{equation}\label{O_T}
O_{\mathrm{T}}\overset{\Delta}{=}\underset{0\leq t\leq T}{\max}\ t-t_{\mathrm{N}}(t).
\vspace{-2mm}\end{equation}
In practice, the maximum outage duration of the UAV usually needs to be designed below a certain value for delay-sensitive communications. For example, if the C$\&$C messages from the GBSs cannot be sent to the UAV reliably (i.e., when outage occurs) for a sufficiently long period, then the UAV may be ``out of control''. In such applications, it is thus critical to design the UAV trajectory $\{{\mv{u}}(t),0\leq t\leq T\}$ such that a maximum outage duration constraint specified by $\bar{O}_{\mathrm{T}}$ can be satisfied, i.e., $O_{\mathrm{T}}\leq \bar{O}_{\mathrm{T}}$.

\vspace{-1mm}
\section{Problem Formulation}\label{sec_formulation}
\vspace{-1mm}
In this paper, we aim to minimize the UAV's mission completion time $T$ by optimizing the UAV trajectory $\{{\mv{u}}(t),0\leq t\leq T\}$, subject to the UAV's initial and final location constraints, as well as the maximum outage duration constraint $O_{\mathrm{T}}\leq \bar{O}_{\mathrm{T}}$. We assume that the UAV flies at its maximum speed $V_{\max}$ (in m/s) during its mission, namely, $\|\dot{\mv{u}}(t)\|=V_{\max}$, where $\dot{\mv{u}}(t)$ denotes the time-derivative of ${\mv{u}}(t)$.\footnote{It can be easily shown that letting the UAV fly at its maximum speed is optimal for our considered problem, which is thus assumed in this paper.} Therefore, we formulate the \hbox{following optimization problem:}
\vspace{-1mm}\begin{align}
\mbox{(P1)} \underset{T,\{{\mv{u}}(t),0\leq t\leq T\}}{\min} & T\label{P1obj}\\[-0.5mm]
\mathrm{s.t.}\quad &{\mv{u}}(0)={\mv{u}}_0\label{P1c1}\\[-0.5mm]
&{\mv{u}}({T})={\mv{u}}_F\label{P1c2}\\[-0.5mm]
&\underset{0\leq t\leq T}{\max}\ t-t_{\mathrm{N}}(t)\leq \bar{O}_{\mathrm{T}} \label{P1c4}\\[-0.5mm]
&\|\dot{\mv{u}}(t)\|= V_{\max},\quad 0\leq t\leq {T}.\label{P1c5}
\end{align}
It is worth noting that for the special case with $\bar{O}_{\mathrm{T}}=0$, i.e., no outage is allowed during the UAV mission, (P1) is equivalent to that studied in our prior work \cite{cellularUAV_arXiv}. Thus, in this paper, we focus on the case of (P1) with $\bar{O}_{\mathrm{T}}> 0$.

Note that (P1) is a non-convex optimization problem, and there are no standard methods to obtain its optimal solution efficiently. Moreover, even checking the feasibility of (P1) for a given $\bar{O}_{\mathrm{T}}> 0$ is a non-trivial problem. To tackle these problems, in the following, we first reformulate (P1) into a more tractable form, based on which we then propose efficient algorithms to check its feasibility and obtain its optimal as well as high-quality suboptimal solutions.

\vspace{-1mm}
\section{Problem Reformulation and Optimal Solution Structure}
\vspace{-1mm}
In this section, we transform (P1) to an equivalent problem in a more tractable form. First, note that one major difficulty in solving (P1) lies in the complicated expression of $t_{\mathrm{N}}(t)$ given in (\ref{t_c}). To tackle this difficulty, we express $t_{\mathrm{N}}(t)$ as well as its associated constraint (\ref{P1c4}) in simplified forms by introducing a so-called \emph{GBS-UAV association sequence}, which specifies a set of GBSs that are successively associated with the UAV to achieve non-outage communications. Then, we show that the optimal trajectory of the UAV should follow a path constituting \emph{connected line segments}, based on which (P1) can be further simplified to jointly design the GBS-UAV association sequence and the corresponding set of \emph{waypoint locations} that specify all line segments in the UAV trajectory.
\vspace{-6mm}
\subsection{GBS-UAV Associations and Problem Reformulation}
\begin{figure}[t]
	\centering
	\includegraphics[width=7cm]{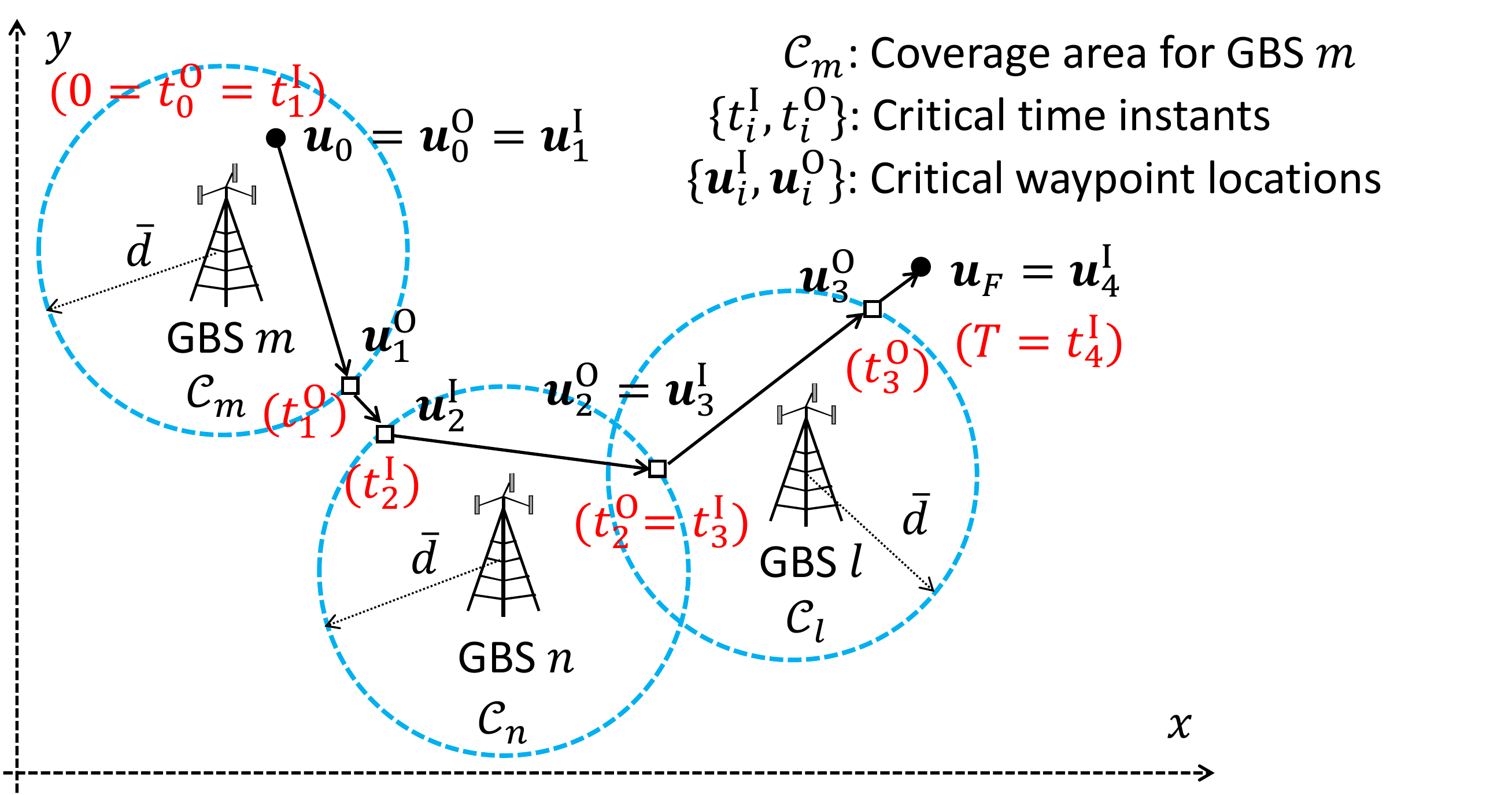}
	\vspace{-3mm}
	\caption{Illustration of GBS-UAV associations with ${\mv{I}}=[m,n,l]^T$.}\label{control}
	\vspace{-6mm}
\end{figure}

\vspace{-1mm}
First, for ease of exposition, we define a so-called \emph{coverage area} for each GBS $m$ as 
\vspace{-1mm}\begin{equation}\label{Cm}
\mathcal{C}_m=\left\{{\mv{u}}\in \mathbb{R}^{2\times 1}: \|{\mv{u}}-{\mv{g}}_m\|\leq \bar{d}\right\},\quad m\in \mathcal{M},
\vspace{-1mm}\end{equation}
which is a disk region on the horizontal plane centered at GBS $m$'s location ${\mv{g}}_m$ with radius $\bar{d}$, as illustrated in Fig. \ref{control}. With (\ref{Cm}), we say that at each time instant $t$, the UAV is \emph{covered by} GBS $m$, i.e., it can be served by GBS $m$ without outage, if its horizontal location lies in $\mathcal{C}_m$, i.e., ${\mv{u}}(t)\in \mathcal{C}_m$. On the other hand, an outage event occurs if the UAV is not covered by any GBS, i.e., ${\mv{u}}(t)\notin \cup_{m\in \mathcal{M}}\mathcal{C}_m$.

Next, we introduce an auxiliary vector ${\mv{I}}=[I_1,...,I_N]^T$ with $I_i\in \mathcal{M},\ \forall i\in \{1,...,N\}$, as the \emph{GBS-UAV association sequence}, which consists of the indices of GBSs that successively associate with the UAV to achieve non-outage communications. Moreover, we introduce a set of auxiliary variables $\{t_i^{\mathrm{I}},t_i^{\mathrm{O}}\}_{i=1}^{N}$ as the \emph{critical time instants}, where $t_i^{\mathrm{I}}$ and $t_i^{\mathrm{O}}$ denote the time instants that the UAV starts to be covered by GBS $I_i$ and stops being covered by it, respectively, with $0\leq t_{1}^{\mathrm{I}}\leq t_{1}^{\mathrm{O}}\leq t_{2}^{\mathrm{I}}\leq ...\leq  t_{N}^{\mathrm{O}}\leq T$. Correspondingly, the UAV is first covered by GBS $I_1$ from $t_1^{\mathrm{I}}$ to $t_1^{\mathrm{O}}$, then by GBS $I_2$ from $t_2^{\mathrm{I}}$ to $t_2^{\mathrm{O}}$, etc., and finally covered by GBS $I_N$ from $t_N^{\mathrm{I}}$ to $t_N^{\mathrm{O}}$. Note that for any $i\in \{2,...,N\}$, if $t_{i-1}^{\mathrm{O}}=t_i^{\mathrm{I}}$ holds, then the UAV is seamlessly handed over from GBS $I_{i-1}$ to GBS $I_i$ without any outage; otherwise, outage occurs during the handover from $t_{i-1}^{\mathrm{O}}$ to $t_i^{\mathrm{I}}$. For convenience, we further define $t_0^{\mathrm{O}}\overset{\Delta}{=}0$ and $t_{N+1}^{\mathrm{I}}\overset{\Delta}{=}T$. In Fig. \ref{control}, we illustrate $\{t_i^{\mathrm{I}},t_i^{\mathrm{O}}\}_{i=1}^{N}$ and $t_0^{\mathrm{O}}$, $t_{N+1}^{\mathrm{I}}$ by taking the example of ${\mv{I}}=[m,n,l]^T$.

With the above definitions, $t_{\mathrm{N}}(t)$ in (\ref{t_c}) {\hbox{can be rewritten as}}
\vspace{-2mm}\begin{equation}\label{tc_eq}
t_{\mathrm{N}}(t)=\begin{cases}
t,\quad &t\in [t_i^{\mathrm{I}},t_i^{\mathrm{O}}),\ \ \ i\in \{1,...,N\}\\
t_{i-1}^{\mathrm{O}},\quad &t\in [t_{i-1}^{\mathrm{O}},t_i^{\mathrm{I}}),\ i\in \{1,...,N+1\}.
\end{cases}
\vspace{-1mm}\end{equation}
The maximum outage duration $O_{\mathrm{T}}$ in (\ref{O_T}) is thus given by
\vspace{-1mm}\begin{equation}\label{O_T_eq}
O_{\mathrm{T}}=\underset{\scriptstyle t\in [t_{i-1}^{\mathrm{O}},t_i^{\mathrm{I}})\atop\scriptstyle i\in\{1,...,N+1\}}{\max}\ t-t_{i-1}^{\mathrm{O}}=\underset{i\in\{1,...,N+1\}}{\max}\ t_{i}^{\mathrm{I}}-t_{i-1}^{\mathrm{O}}.
\vspace{-2mm}\end{equation}
Then, we are ready to present the following proposition.
\begin{proposition}\label{prop_P1eq}
	(P1) is equivalent to the following problem:
		\vspace{-5mm}
		\begin{align}
	\mbox{(P2)} \underset{\scriptstyle {\mv{I}},\{t_i^{\mathrm{I}},t_i^{\mathrm{O}}\}_{i=1}^N
		\atop \scriptstyle T,\{{\mv{u}}(t),0\leq t\leq {T}\}}{\min}\ & T \label{P2obj}\\[-0.5mm]
	\mathrm{s.t.}\quad & (\ref{P1c1}),(\ref{P1c2}), (\ref{P1c5})\\[-0.5mm]
	&\|{\mv{u}}({t})-{\mv{g}}_{I_i}\|\leq \bar{d},\quad t_i^{\mathrm{I}}\leq t\leq t_i^{\mathrm{O}},\nonumber\\&\qquad\qquad\qquad\qquad\ \ i=1,...,N\label{P2c1}\\[-0.5mm]
	& t_0^{\mathrm{O}}=0\label{P2c2}\\[-0.5mm]
	& t_{N+1}^{\mathrm{I}}=T\label{P2c3}\\[-0.5mm]
	& t_{i-1}^{\mathrm{O}}\leq t_{i}^{\mathrm{I}}\leq t_{i}^{\mathrm{O}}\leq T,\ i=1,...,N\label{P2c4}\\
	& I_i\in \mathcal{M},\quad i=1,...,N\label{P2c5}\\[-0.5mm]
	& t_{i}^{\mathrm{I}}-t_{i-1}^{\mathrm{O}}\leq \bar{O}_{\mathrm{T}},\ i=1,...,N+1.\label{P2c6}
	\end{align}
\end{proposition}
\vspace{-1mm}
\begin{IEEEproof}
	Please refer to Appendix \ref{proof_prop_P1eq}.
\end{IEEEproof}

Denote the horizontal location of the UAV at the critical time instants $t_i^{\mathrm{I}}$ and $t_i^{\mathrm{O}}$ as ${\mv{u}}_{i}^{\mathrm{I}}\overset{\Delta}{=}{\mv{u}}(t_i^{\mathrm{I}})$ and ${\mv{u}}_{i}^{\mathrm{O}}\overset{\Delta}{=}{\mv{u}}(t_i^{\mathrm{O}})$, respectively. Note that since the UAV is covered by GBS $I_i$ at both $t_i^{\mathrm{I}}$ and $t_i^{\mathrm{O}}$, we have
${\mv{u}}_{i}^{\mathrm{I}}\in \mathcal{C}_{I_i},\ {\mv{u}}_{i}^{\mathrm{O}}\in \mathcal{C}_{I_i},\ i=1,...,N$. In the following, we refer to $\{{\mv{u}}_i^{\mathrm{I}},{\mv{u}}_i^{\mathrm{O}}\}_{i=1}^N$ as the set of {\emph{critical waypoint locations}}, as also illustrated in Fig. \ref{control} for the example of ${\mv{I}}=[m,n,l]^T$. In addition, note that we also have ${\mv{u}}_0^{\mathrm{O}}\overset{\Delta}{=}{\mv{u}}_0$ and ${\mv{u}}_{N+1}^{\mathrm{I}}\overset{\Delta}{=}{\mv{u}}_F$ by definition.

\vspace{-2mm}
\subsection{Optimal Structure of UAV Trajectory}
\vspace{-1mm}
Based on the reformulated problem (P2), we show a simplified structure of the optimal UAV trajectory.
\begin{proposition}[Trajectory with Connected Line Segments]\label{prop_connected}
	The optimal solution to (P2) satisfies the following conditions:
	\begin{align}
	t_i^{\mathrm{I}}&=t_{i-1}^{\mathrm{O}}+\|{\mv{u}}_i^{\mathrm{I}}-{\mv{u}}_{i-1}^{\mathrm{O}}\|/V_{\max},\quad i=1,...,N+1,\label{tiI}\\[-0.5mm]
	t_i^{\mathrm{O}}&=t_i^{\mathrm{I}}+\|{\mv{u}}_i^{\mathrm{O}}-{\mv{u}}_{i}^{\mathrm{I}}\|/V_{\max},\quad\quad\ \ i=1,...,N,\label{tiO}\\[-0.5mm]
	{\mv{u}}(t)\!&=\!\begin{cases}
	{\mv{u}}_{i-1}^{\mathrm{O}}\!+\!(t-t_{i-1}^{\mathrm{O}})V_{\max}\frac{{\mv{u}}_{i}^{\mathrm{I}}-{\mv{u}}_{i-1}^{\mathrm{O}}}{\|{\mv{u}}_{i}^{\mathrm{I}}-{\mv{u}}_{i-1}^{\mathrm{O}}\|},t\in [t_{i-1}^{\mathrm{O}},t_{i}^{\mathrm{I}}],\\[-0.5mm]
	\qquad \qquad \qquad \qquad \qquad \qquad \qquad i=1,...,N+1\\[-0.5mm]
	{\mv{u}}_{i}^{\mathrm{I}}\!+\!(t-t_{i}^{\mathrm{I}})V_{\max}\frac{{\mv{u}}_{i}^{\mathrm{O}}-{\mv{u}}_{i}^{\mathrm{I}}}{\|{\mv{u}}_{i}^{\mathrm{O}}-{\mv{u}}_{i}^{\mathrm{I}}\|},t\!\in\! [t_{i}^{\mathrm{I}},t_{i}^{\mathrm{O}}],i\!=\!1,...,N,
	\end{cases}\\[-2mm]
	T&=\sum_{i=1}^{N+1}\|{\mv{u}}_{i}^{\mathrm{I}}\!-\!{\mv{u}}_{i-1}^{\mathrm{O}}\|/V_{\max}\!+\!\sum_{i=1}^N \|{\mv{u}}_{i}^{\mathrm{O}}\!-\!{\mv{u}}_{i}^{\mathrm{I}}\|/V_{\max}.\label{T}
	\end{align}
\end{proposition}
	\vspace{-2mm}
	\begin{IEEEproof}
	Please refer to Appendix \ref{proof_prop_connected}.
\end{IEEEproof}

Note that according to (\ref{tiI})--(\ref{T}), the UAV should fly from $U_0$ to $U_F$ following a path consisting of \emph{connected line segments} only, where the end points that determine these connected line segments are the critical waypoints with horizontal locations $\{{\mv{u}}_0,{\mv{u}}_1^{\mathrm{I}},{\mv{u}}_1^{\mathrm{O}},{\mv{u}}_2^{\mathrm{I}},{\mv{u}}_2^{\mathrm{O}},..., {\mv{u}}_N^{\mathrm{I}},{\mv{u}}_N^{\mathrm{O}},{\mv{u}}_F\}$. Hence, it can be shown that (P2) is equivalent to the following problem based on Proposition \ref{prop_connected}:
\vspace{-2mm}\begin{align}
\mbox{(P3)} \underset{{\mv{I}},\{{\mv{u}}_i^{\mathrm{I}},{\mv{u}}_i^{\mathrm{O}}\}_{i=1}^N
}{\min} &\sum_{i=1}^{N+1}\|{\mv{u}}_i^{\mathrm{I}}-{\mv{u}}_{i-1}^{\mathrm{O}}\|+\sum_{i=1}^N\|{\mv{u}}_i^{\mathrm{O}}-{\mv{u}}_i^{\mathrm{I}}\|\\[-0.5mm]
\mathrm{s.t.}\quad &{\mv{u}}_0^{\mathrm{O}}={\mv{u}}_0\label{P3c1}\\[-0.5mm]
& {\mv{u}}_{N+1}^{\mathrm{I}}={\mv{u}}_F\label{P3c2}\\[-0.5mm]
& \|{\mv{u}}_i^{\mathrm{I}}-{\mv{g}}_{I_i}\|\leq \bar{d},\quad i=1,...,N\label{P3c3}\\[-0.5mm]
& \|{\mv{u}}_i^{\mathrm{O}}-{\mv{g}}_{I_i}\|\leq \bar{d},\quad i=1,...,N\label{P3c4}\\[-0.5mm]
& I_i\in\mathcal{M},\quad i=1,...,N\label{P3c5}\\[-0.5mm]
& I_i\neq I_j,\quad i\neq j,\ i,j=1,...,N\label{P3c6}\\[-0.5mm]
& \underset{i\in \{1,...,N+1\}}{\max}\|{\mv{u}}_{i}^{\mathrm{I}}-{\mv{u}}_{i-1}^{\mathrm{O}}\|\leq V_{\max}\bar{O}_{\mathrm{T}}.\label{P3c7}
\end{align}
\vspace{-4mm}

Notice that Problem (P3) is a joint optimization problem for the GBS-UAV association sequence $\mv{I}$ and the corresponding waypoint locations $\{{\mv{u}}_i^{\mathrm{I}},{\mv{u}}_i^{\mathrm{O}}\}_{i=1}^N$. Furthermore, (P3) is equivalent to (P1), but (P3) involves a significantly reduced number of variables as compared to (P1), thanks to the optimal line-segment structure of the UAV trajectory. Note that (P1) is feasible if and only if (P3) is feasible, thus the feasibility of (P1) can be equivalently verified by checking the feasibility of (P3); moreover, the optimal solution to (P1) can be obtained by substituting the optimal solution obtained for (P3) into (\ref{tiI})--(\ref{T}). Therefore, we focus on solving (P3) in the next.
\vspace{-2mm}
\section{Feasibility Check}\label{sec_feas}
\vspace{-2mm}
Prior to solving Problem (P3), we check its feasibility in this section. Note that (P3) is feasible if and only if the problem below is feasible, and its optimal value is no larger than $\bar{O}_{\mathrm{T}}$:
\vspace{-1mm}\begin{align}
\mbox{(P3-F)} \underset{{\mv{I}},\{{\mv{u}}_i^{\mathrm{I}},{\mv{u}}_i^{\mathrm{O}}\}_{i=1}^N
}{\min} &\underset{i\in \{1,...,N+1\}}{\max}\|{\mv{u}}_{i}^{\mathrm{I}}-{\mv{u}}_{i-1}^{\mathrm{O}}\|/V_{\max}\\[-0.5mm]
\mathrm{s.t.}\quad &(\ref{P3c1}),(\ref{P3c2}),(\ref{P3c3}),(\ref{P3c4}),(\ref{P3c5}),(\ref{P3c6}).
\end{align}
It is worth noting that the optimal value of (P3-F) represents the minimum achievable maximum outage duration. For (P3-F), we have the following lemma.
\begin{lemma}[Maximum Outage Duration Minimizing Waypoint Locations]\label{prop_waypoint}
Given any $\mv{I}$ that satisfies the constraints in (\ref{P3c5}) and (\ref{P3c6}), the maximum outage duration is minimized as
\vspace{-1mm}
\begin{align}\label{O_TI}
O_{\mathrm{T}}^\star({\mv{I}})=\max&\big\{\|{\mv{u}}_0-{\mv{g}}_{I_1}\|-\bar{d},\|{\mv{u}}_F-{\mv{g}}_{I_N}\|-\bar{d},\nonumber\\[-0.5mm]
&\underset{i\in\{2,...,N\}}{\max}\|{\mv{g}}_{I_{i}}-{\mv{g}}_{I_{i-1}}\|-2\bar{d},0\big\}/V_{\max}.
\end{align}
\vspace{-1mm}
An optimal solution of $\{{\mv{u}}_i^{\mathrm{I}},{\mv{u}}_i^{\mathrm{O}}\}_{i=1}^N$ to (P3-F) is given by
\vspace{-1mm}
\begin{align}
{\mv{u}}_{1}^{\mathrm{I}}&\!=\!\begin{cases}{\mv{u}}_0,\qquad\qquad\qquad\qquad\quad {\mathrm{if}}\ \|{\mv{u}}_0\!-\!{\mv{g}}_{I_1}\|\leq \bar{d}\\[-0.5mm] {\mv{g}}_{I_1}\!-\!\bar{d}({\mv{g}}_{I_1}\!-\!{\mv{u}}_0)/\|{\mv{g}}_{I_1}\!-\!{\mv{u}}_0\|,\quad  \mathrm{otherwise},\end{cases}\!\!\!\!\!\!\!\label{P3Fu1}\\[-0.5mm]
{\mv{u}}_{N}^{\mathrm{O}}&\!=\!\begin{cases}{\mv{u}}_F,\qquad\qquad\qquad\qquad\ {\mathrm{if}}\ \|{\mv{u}}_F\!-\!{\mv{g}}_{I_N}\|\leq \bar{d}\\[-0.5mm] 
{\mv{g}}_{I_N}\!+\!\bar{d}({\mv{u}}_F\!-\!{\mv{g}}_{I_N})/\|{\mv{u}}_F\!-\!{\mv{g}}_{I_N}\|,\mathrm{otherwise},\end{cases}\!\!\!\label{P3Fu2}\\[-0.5mm]
{\mv{u}}_{i-1}^{\mathrm{O}}&\!=\!{\mv{g}}_{I_{i-1}}\!\!+\!\bar{d}({\mv{g}}_{I_{i}}\!-\!{\mv{g}}_{I_{i-1}})/\|{\mv{g}}_{I_i}\!-\!{\mv{g}}_{I_{i-1}}\|, i\!=\!2,...,N,\!\!\label{P3Fu3}\\[-0.5mm]
{\mv{u}}_i^{\mathrm{I}}&\!=\!\begin{cases}
{\mv{u}}_{i-1}^{\mathrm{O}},\quad\qquad\qquad\qquad \mathrm{if}\ \|{\mv{g}}_{I_i}-{\mv{g}}_{I_{i-1}}\|\leq 2\bar{d}\\[-0.5mm]
{\mv{g}}_{I_i}\!\!-\!\bar{d}({\mv{g}}_{I_i}-{\mv{g}}_{I_{i-1}})/\|{\mv{g}}_{I_i}-{\mv{g}}_{I_{i-1}}\|, \mathrm{otherwise},\end{cases}\nonumber\\[-0.5mm]
&\qquad\qquad\qquad\qquad\qquad\qquad\qquad\quad i=2,...,N.\label{P3Fu4}
\end{align}
\vspace{-6mm}
\end{lemma}
\begin{IEEEproof}
	Please refer to Appendix \ref{proof_prop_waypoint}.
\end{IEEEproof}
\begin{figure}[t]
	\centering
	\includegraphics[width=7cm]{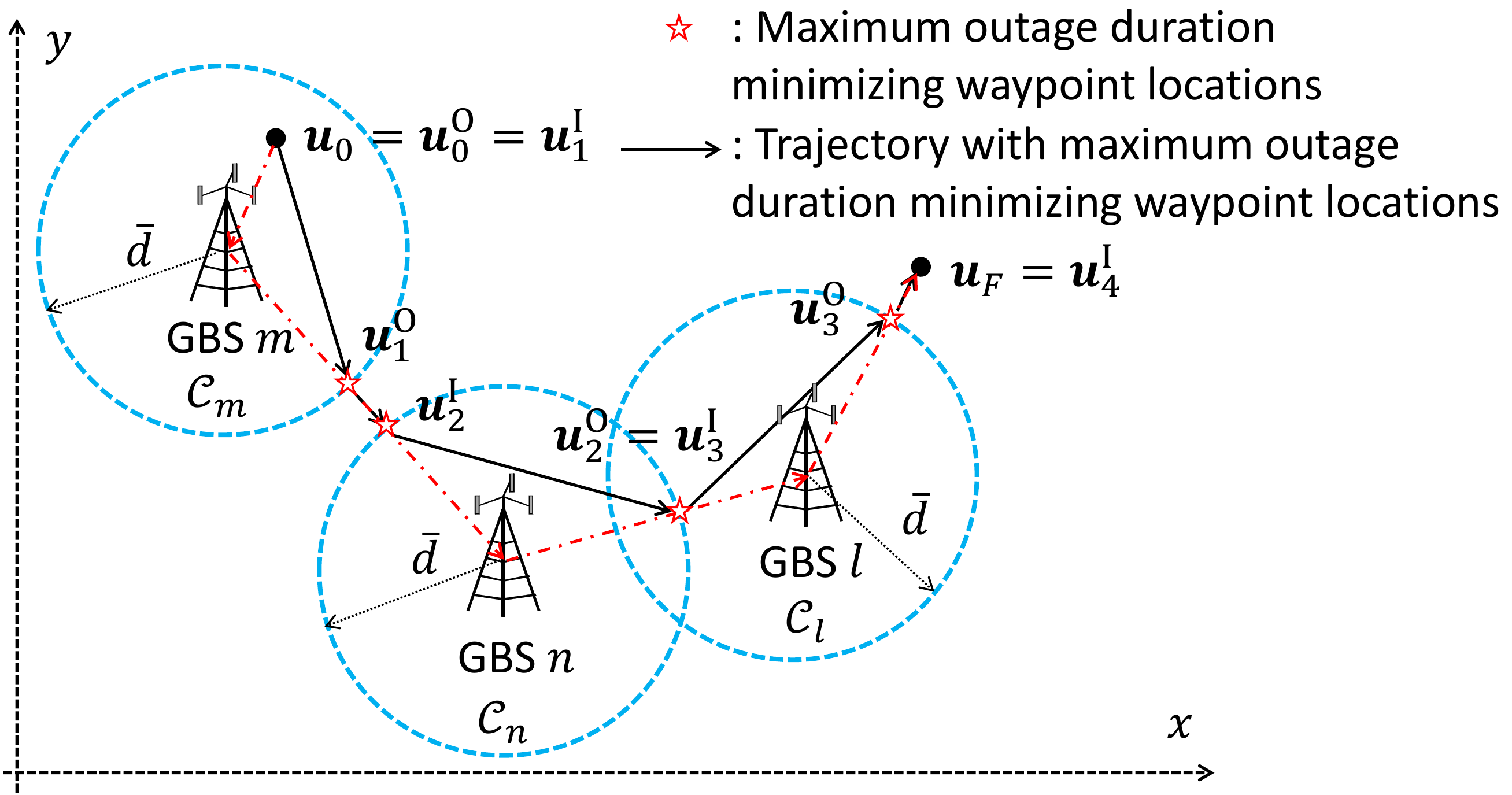}
	\vspace{-3mm}
	\caption{Illustration of maximum outage duration minimizing waypoint locations with ${\mv{I}}=[m,n,l]^T$.}\label{minReliability}
	\vspace{-6mm}
\end{figure}
Note that Lemma \ref{prop_waypoint} suggests that the maximum outage duration is minimized by placing the waypoint locations ${\mv{u}}_i^{\mathrm{I}}$ and ${\mv{u}}_{i-1}^{\mathrm{O}}$ on the boundaries of $\mathcal{C}_{I_i}$ or $\mathcal{C}_{I_{i-1}}$ and at the same time on the connected line segment between GBSs $I_{i-1}$ and $I_i$, as given in (\ref{P3Fu1})--(\ref{P3Fu4}) and illustrated in Fig. \ref{minReliability}. As a direct result of Lemma \ref{prop_waypoint}, we have the following proposition, for which the proof is omitted for brevity.
\begin{proposition}\label{prop_feas}
	(P3) is feasible if and only if there exists an $\mv{I}$ which satisfies (\ref{P3c5}), (\ref{P3c6}) and the following conditions:
	\vspace{-1mm}\begin{align}
	\|{\mv{u}}_0-{\mv{g}}_{I_1}\|-\bar{d}&\leq V_{\max}\bar{O}_{\mathrm{T}},\\[-0.5mm] \|{\mv{g}}_{I_i}-{\mv{g}}_{I_{i-1}}\|-2\bar{d}&\leq V_{\max}\bar{O}_{\mathrm{T}},\quad i=2,...,N,\\[-0.5mm]
	\|{\mv{u}}_F-{\mv{g}}_{I_N}\|-\bar{d}&\leq V_{\max}\bar{O}_{\mathrm{T}}.
	\end{align}
\end{proposition}

\vspace{-2mm}
Based on Proposition \ref{prop_feas}, we provide a \emph{graph theory} based approach to check the feasibility of (P3). Specifically, we construct an undirected graph denoted by $G_{\mathrm{M}}=(V_{\mathrm{M}},E_{\mathrm{M}})$ \cite{graph}. The vertex set of $G_{\mathrm{M}}$ is given by
\vspace{-2mm}\begin{equation}
V_{\mathrm{M}}=\{U_0,G_1,G_2,...,G_M,U_F\}.
\vspace{-1mm}\end{equation}
The edge set of $G_{\mathrm{M}}$ is given by
\vspace{-2mm}\begin{align}
&E_{\mathrm{M}}=\{(U_0,G_m):\|{\mv{u}}_0-{\mv{g}}_m\|-\bar{d}\leq V_{\max}\bar{O}_{\mathrm{T}},m\in \mathcal{M}\}\cup\nonumber\\[-0.5mm]
&\{(G_m,G_n)\!:\!\|{\mv{g}}_m\!-\!{\mv{g}}_n\|\!-\!2\bar{d}\leq V_{\max}\bar{O}_{\mathrm{T}},m,n\!\in\!\mathcal{M},
m\neq n\}\nonumber\\[-0.5mm]
&\cup\{(U_F,G_m):\|{\mv{u}}_F-{\mv{g}}_m\|-\bar{d}\leq V_{\max}\bar{O}_{\mathrm{T}}, m\in \mathcal{M}\}.\!
\end{align}

\vspace{-2mm}
Note that in $G_{\mathrm{M}}$, an edge $(U_0,G_m)$ exists if and only if the minimum outage duration from the mission start time to the instant that the UAV starts to be covered by GBS $m$ is no larger than $\bar{O}_{\mathrm{T}}$; an edge $(U_F,G_m)$ exists if and only if the minimum outage duration from the instant that the UAV stops being covered by GBS $m$ to the mission completion time is no larger than $\bar{O}_{\mathrm{T}}$; and an edge $(G_m,G_n)$ exists if and only if the minimum outage duration from the instant that the UAV stops being covered by GBS $m$ to that the UAV starts to be covered by GBS $n$ is no larger than $\bar{O}_{\mathrm{T}}$. Therefore, it follows from Proposition \ref{prop_feas} that Problem (P3) is feasible if and only $U_0$ and $U_F$ in $G_{\mathrm{M}}$ are \emph{connected} \cite{graph}. Hence, the feasibility of (P3) can be checked by constructing $G_{\mathrm{M}}$ with complexity of $\mathcal{O}(M^2)$, and checking the connectivity between $U_0$ and $U_F$ in $G_{\mathrm{M}}$ via e.g., breadth-first search with complexity of $\mathcal{O}(M^2)$, thus requiring an overall complexity of $\mathcal{O}(M^2)$ \cite{graph}. 

\vspace{-1mm}
\section{Proposed Solution to (P3)}
\vspace{-1mm}
In this section, we solve Problem (P3) assuming that it has been verified to be feasible. Note that (P3) is a non-convex combinatorial optimization problem due to the discrete variables in $\mv{I}$, where the length of $\mv{I}$, $N$, is also an implicit variable. Moreover, ${\mv{I}}$ and  $\{{\mv{u}}_i^{\mathrm{I}},{\mv{u}}_i^{\mathrm{O}}\}$ are coupled by the constraints in (\ref{P3c3}), (\ref{P3c4}) and (\ref{P3c7}), which makes the problem more difficult to solve. In the following, we apply graph theory and convex optimization techniques to overcome the above challenges, and propose the optimal solution as well as a lower-complexity suboptimal solution to (P3), respectively. 
\vspace{-1mm}
\subsection{Optimal Solution}
\vspace{-1mm}
Note that with any given GBS-UAV association sequence $\mv{I}$ that satisfies (\ref{P3c5}) and (\ref{P3c6}), (P3) is a convex optimization problem over the waypoint locations $\{{\mv{u}}_i^{\mathrm{I}},{\mv{u}}_i^{\mathrm{O}}\}_{i=1}^N$, which can be efficiently solved via existing software, e.g., CVX \cite{cvx}, with polynomial complexity over $N$, e.g., $\mathcal{O}(N^{3.5})$ by casting this problem as a second-order cone program (SOCP) \cite{convex}. Moreover, recall that each feasible solution of $\mv{I}$ to (P3) corresponds to a \emph{path} between $U_0$ and $U_F$ in graph $G_{\mathrm{M}}$ constructed in the preceding section, where all such paths can be found via existing algorithms in graph theory, e.g., the depth-first search method with complexity $\mathcal{O}(M!)$ \cite{graph}. Hence, the optimal solution to (P3) can be obtained by finding all such paths as well as the corresponding optimal $\{{\mv{u}}_i^{\mathrm{I}},{\mv{u}}_i^{\mathrm{O}}\}_{i=1}^N$'s, and selecting the one with the minimum objective value, which requires a worst-case complexity of $\mathcal{O}(M^{3.5}M!)$ since $N\leq M$ holds due to the constraints in (\ref{P3c6}). 
\vspace{-1mm}
\subsection{Suboptimal Solution}
\vspace{-1mm}
To further reduce the complexity of the optimal solution, especially when the number of involved GBSs, $M$, is practically large (e.g., when the initial and final locations of the UAV, $U_0$ and $U_F$, are far apart), we propose an alternative approach for finding an approximate solution to (P3). Specifically, we find an approximate solution of the GBS-UAV association sequence $\mv{I}$ firstly, and then obtain the optimal waypoint locations $\{{\mv{u}}_i^{\mathrm{I}},{\mv{u}}_i^{\mathrm{O}}\}_{i=1}^N$ with the obtained $\mv{I}$ via CVX (similarly as in the optimal solution). Thus, our remaining task is to find an approximate solution of $\mv{I}$, for which we present a new \emph{graph} based problem reformulation of (P3) by applying appropriate bounding \hbox{and approximation techniques as follows.}

Recall from Lemma \ref{prop_waypoint} that (P3) is feasible with given $\mv{I}$ if and only if the waypoint locations $\{{\mv{u}}_i^{\mathrm{I}},{\mv{u}}_i^{\mathrm{O}}\}_{i=1}^N$ given in (\ref{P3Fu1})--(\ref{P3Fu4}) satisfy the constraint in (\ref{P3c7}). Therefore, we can find an approximate solution of $\mv{I}$ by substituting (\ref{P3Fu1})--(\ref{P3Fu4}) into (P3). Nevertheless, note that it is generally difficult to explicitly express the objective function of (P3) with given $\mv{I}$ and the corresponding waypoints in (\ref{P3Fu1})--(\ref{P3Fu4}). Thus, we further consider an upper bound of the objective value of (P3), denoted by $s_D$, which is given by
\vspace{-2mm}\begin{equation}\label{sDbound}
\!\!s_D\!\leq\! \|{\mv{u}}_0\!-\!{\mv{g}}_{I_1}\|\!+\!\sum_{i=2}^N\|{\mv{g}}_{I_i}\!-\!{\mv{g}}_{I_{i-1}}\|\!+\!\|{\mv{u}}_F\!-\!{\mv{g}}_{I_N}\|\!\overset{\Delta}{=}\!\bar{s}_D.
\vspace{-2mm}\end{equation}
Note that (\ref{sDbound}) can be proved via the triangle inequality and is illustrated in Fig. \ref{minReliability}. Hence, to find an approximate solution of $\mv{I}$, we solve (P3) by replacing its objective function by $\bar{s}_D$ given in (\ref{sDbound}), and considering the additional constraints in (\ref{P3Fu1})--(\ref{P3Fu4}), for which we provide a graph based solution below.

Consider the same graph $G_{\mathrm{M}}=(V_{\mathrm{M}},E_{\mathrm{M}})$ as constructed in Section \ref{sec_feas}. We further consider a  set of weights for $G_{\mathrm{M}}$:
\vspace{-1mm}
\begin{align}\label{W_M}
&W_{\mathrm{M}}(U_0,G_m)=\|{\mv{u}}_0-{\mv{g}}_m\|,\ W_{\mathrm{M}}(U_F,G_m)=\|{\mv{u}}_F-{\mv{g}}_m\|\nonumber\\[-1mm]
&W_{\mathrm{M}}(G_m,G_n)=\|{\mv{g}}_m-{\mv{g}}_n\|,\quad m,n\in \mathcal{M},m\neq n.
\end{align}
With the constructed $G_{\mathrm{M}}$, the aforementioned problem is equivalent to finding the \emph{shortest path} from $U_0$ to $U_F$ in $G_{\mathrm{M}}$ with respect to the weights $W_{\mathrm{M}}$'s, which can be efficiently solved via e.g., the Dijkstra algorithm with complexity of $\mathcal{O}(M^2)$ \cite{graph}. By further noting that constructing the graph $G_{\mathrm{M}}$ also requires complexity of $\mathcal{O}(M^2)$, the overall complexity for finding a suboptimal solution of $\mv{I}$ is thus $\mathcal{O}(M^2)$. Note that the worst-case complexity for obtaining the optimal waypoint locations with the given $\mv{I}$ is $\mathcal{O}(M^{3.5})$. Hence, the overall complexity for finding a suboptimal solution to (P3) is $\mathcal{O}(M^{3.5})$, which is significantly reduced as compared to $\mathcal{O}(M^{3.5}M!)$ for finding the optimal solution when $M$ is large.
\vspace{-2mm}
\section{Numerical Examples}\label{sec_numerical}
\begin{figure}[b]
	\vspace{-8mm}	
	\centering
	\includegraphics[width=8.5cm]{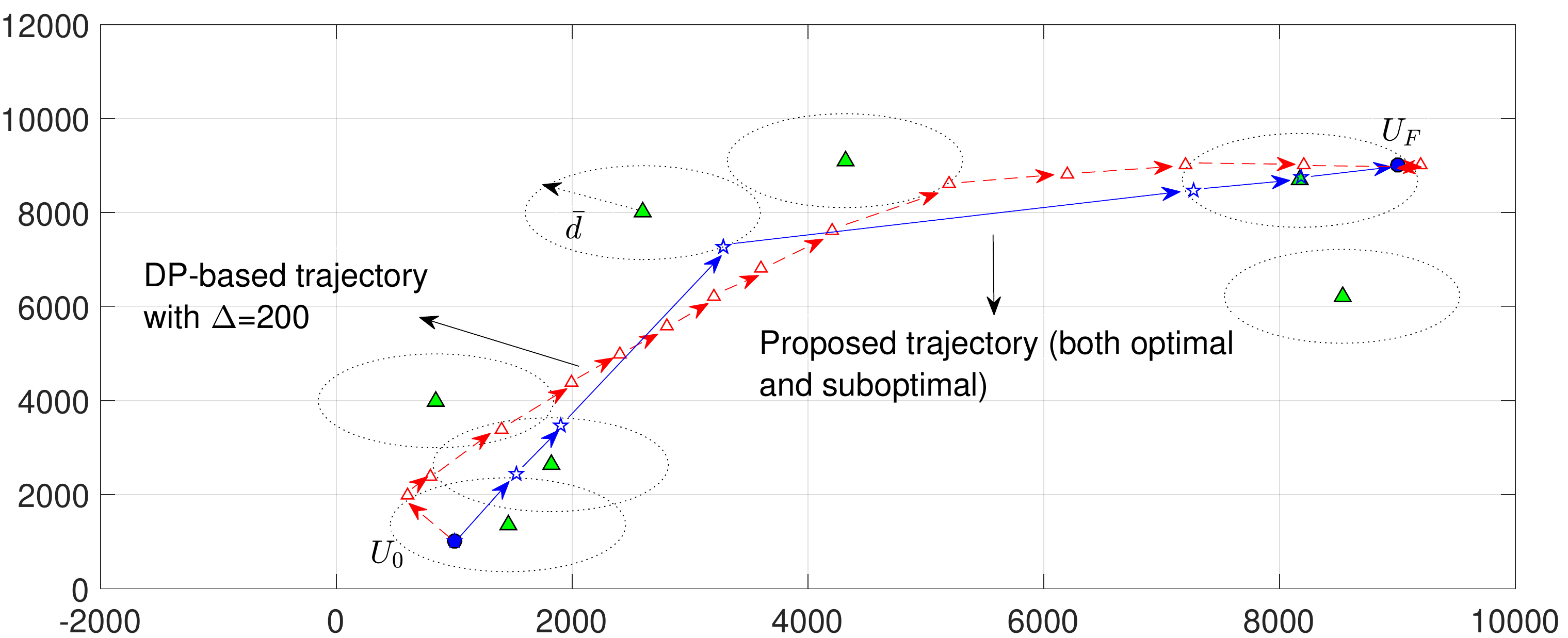}
	\vspace{-3mm}
	\caption{Trajectory comparison with $\bar{O}_{\mathrm{T}}=84.6124$ s.}\label{trajectory}
\end{figure}

\vspace{-1mm}
In this section, we provide numerical examples. We consider $M=7$ GBSs that are randomly distributed in a $D\ \mathrm{km}\times D\ \mathrm{km}$ region, with $D=10$. One random realization of GBSs' locations is shown in Fig. \ref{trajectory}. The UAV's initial and final locations projected on the horizontal plane are set as ${\mv{u}}_0=[1000,1000]^T$ and ${\mv{u}}_F=[9000,9000]^T$, respectively. The altitude of the UAV and each GBS are set as $H=90$ m and $H_{\mathrm{G}}=12.5$ m, respectively. The maximum UAV speed is set as $V_{\max}=50$ m/s. The reference SNR at distance $d_0=1$ m is set as $\rho_0=\frac{P\beta_0}{\sigma^2}=80$ dB. The minimum received SNR target is set as $\bar{\rho}=20$ dB.

\begin{figure}[t]
	\centering
	\includegraphics[width=8.5cm]{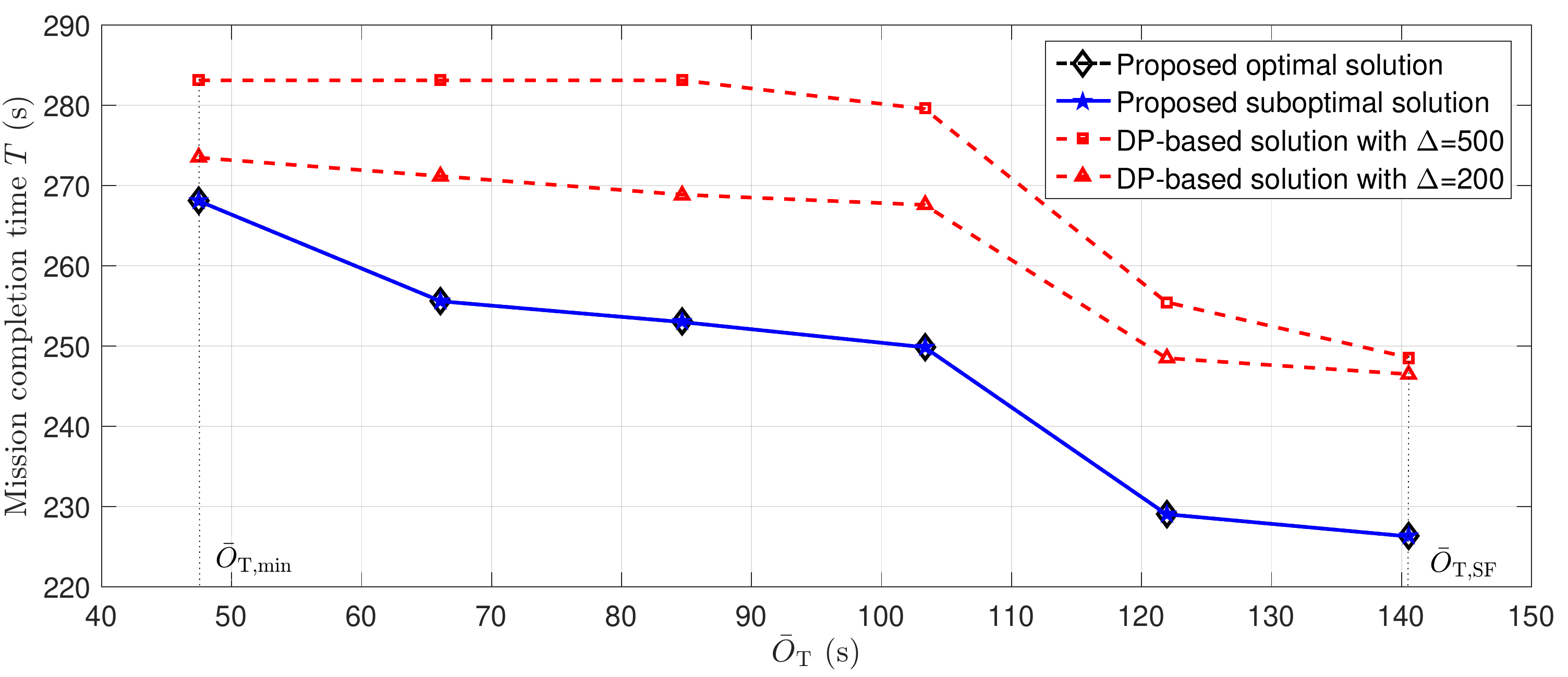}
	\vspace{-4mm}
	\caption{Mission completion time $T$ versus $\bar{O}_{\mathrm{T}}$.}\label{mission}
	\vspace{-7mm}
\end{figure}
Under the above setup, we first minimize the maximum outage duration via the bi-section method based on Proposition \ref{prop_feas}, which is obtained as $\bar{O}_{\mathrm{T},\min}=47.4941$ s. Then, we obtain an upper bound on the maximum outage duration with the straight-flight (SF) trajectory from $U_0$ to $U_F$, which is \hbox{$\bar{O}_{\mathrm{T,SF}}=140.5163$ s.} Note that the SF trajectory achieves the minimum mission completion time $T$, thus the optimal solution to (P1) with any $\bar{O}_{\mathrm{T}}>\bar{O}_{\mathrm{T,SF}}$ can be easily shown to be the \hbox{SF trajectory.} Hence, we evaluate our proposed trajectory designs for $\bar{O}_{\mathrm{T}}\in[\bar{O}_{\mathrm{T},\min},\bar{O}_{\mathrm{T,SF}}]$ in the following.

\begin{figure}[t]
	\centering
	\includegraphics[width=8.5cm]{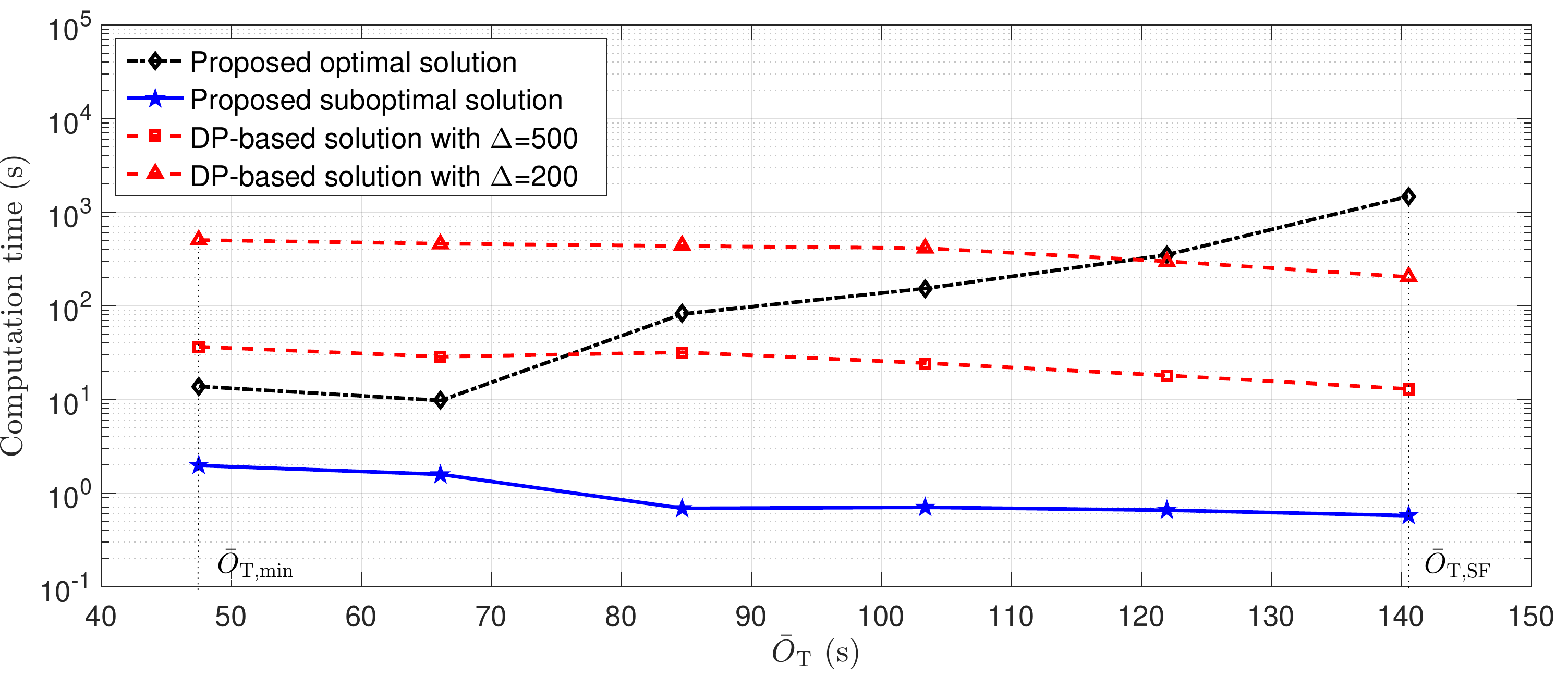}
	\vspace{-4mm}
	\caption{Computation time versus $\bar{O}_{\mathrm{T}}$.}\label{CPU}
	\vspace{-6mm}
\end{figure}

For comparison, we consider the DP-based trajectory design proposed in \cite{Disconnectivity} as a benchmark scheme. In the DP-based design, the $D\ \mathrm{km}\times D\ \mathrm{km}$ area is quantized to a grid with granularity $\Delta$ m, which specifies the set of possible UAV locations during its flight. The trajectory is then found in a recursive manner from $U_F$ to $U_0$ by iteratively updating the ``best'' last grid point (from a set of neighboring points within a $2n_r\times 2n_r$ square) before the UAV arrives at each grid point via exploiting the sub-problem structure \cite{Disconnectivity}. We set $\Delta=200$ or $500$ m, and $n_r=1000$ m. In Fig. \ref{mission}, we show the mission completion time $T$ versus the maximum outage duration target $\bar{O}_{\mathrm{T}}$ for the proposed optimal and suboptimal solutions as compared to the DP-based solution with $\Delta=200$ or $500$. It is observed that our proposed suboptimal solution achieves the same performance as the optimal solution, thus validating the efficacy of the bounding and approximation techniques applied for solving (P3) shown in Section VI-B. 
Moreover, it is observed that the DP-based solution with $\Delta=200$ outperforms that with $\Delta=500$, since smaller granularity results in finer-grained UAV locations and thus better performance. Furthermore, both DP-based solutions are outperformed by our proposed solutions. This is because with the optimal trajectory structure given in Proposition \ref{prop_connected}, our proposed solutions only need to find critical parameters such as the GBS-UAV association sequence and the waypoint locations, thus being more efficient than the DP-based solutions which require quantization of the entire area of interest to check all possible UAV locations. Furthermore, we consider $\bar{O}_{\mathrm{T}}=84.6124$ s (the third sample of $\bar{O}_{\mathrm{T}}$ in Fig. \ref{mission}) and illustrate in Fig. \ref{trajectory} the proposed and DP-based (with $\Delta=200$) trajectories, which are observed to be substantially different. Last, we show in Fig. \ref{CPU} the required computation time for the different trajectory design solutions.\footnote{All the computations are executed by MATLAB on a computer with an Intel Core i5 $3.40$-GHz CPU and $8$ GB of memory.} It is observed that the computation time for DP-based solutions increases as $\Delta$ decreases, due to the rapidly enlarged state-space set. In contrast, our proposed suboptimal solution requires much less computation time than the DP-based solutions as well as the proposed optimal solution; thus it is a practically appealing solution from both performance and complexity considerations.

\vspace{-2mm}
\section{Conclusion}\label{sec_conclusion}
\vspace{-1mm}
This paper studies the trajectory design for cellular-connected UAVs under delay-limited communications. We consider a minimum received SNR target for non-outage UAV communications, based on which the UAV trajectory is optimized to minimize the UAV's mission completion time from an initial location to a final location, subject to a constraint on the maximum tolerable outage duration in the flight. By exploiting the optimal structure of the trajectory solution, we apply graph theory and convex optimization to devise efficient algorithms to check the problem feasibility and find both optimal and low-complexity suboptimal solutions. Numerical examples validate the efficacy of our proposed designs.

\appendix
\subsection{Proof of Proposition \ref{prop_P1eq}}\label{proof_prop_P1eq}
First, for any feasible solution $(\tilde{T},\{\tilde{\mv{u}}(t),0\!\leq \!\! t\!\leq \!\tilde{T}\})$ to (P1), we can always construct a GBS-UAV association sequence $\tilde{\mv{I}}=[\tilde{I}_1,...,\tilde{I}_{\tilde{N}}]^{\tilde{N}}$ and a set of critical time instants $\{\tilde{t}_i^{\mathrm{I}},\tilde{t}_{i}^{\mathrm{O}}\}_{i=1}^{\tilde{N}}$ based on their definitions presented in Section IV, which satisfy the constraints in (\ref{P2c1})--(\ref{P2c5}). Moreover, it follows from (\ref{tc_eq}) and (\ref{O_T_eq}) that $\!\!\underset{i\in\{1,...,\tilde{N}+1\}}{\max}\!\!\tilde{t}_{i}^{\mathrm{I}}\!-\!\tilde{t}_{i-1}^{\mathrm{O}}\!=\!\underset{0\leq t\leq \tilde{T}}{\max}t\!-\!t_{\mathrm{N}}(t)\!\leq \!\bar{O}_{\mathrm{T}}$ holds, thus $(\tilde{\mv{I}},\{\tilde{t}_i^{\mathrm{I}},\tilde{t}_i^{\mathrm{O}}\}_{i=1}^{\tilde{N}},  \tilde{T},\{\tilde{\mv{u}}(t),0\!\leq \!t\!\leq \!\tilde{T}\})$ is a feasible solution to (P2) with the same objective value as (P1) with the solution $(\tilde{T},\{\tilde{\mv{u}}(t),0\!\!\leq \!\!t\!\!\leq \!\tilde{T}\})$. Hence, the optimal value of (P2) is no larger than that of (P1).
On the other hand, for any feasible solution $(\tilde{\mv{I}},\{\tilde{t}_i^{\mathrm{I}},\tilde{t}_i^{\mathrm{O}}\}_{i=1}^{\tilde{N}},  \tilde{T},\{\tilde{\mv{u}}(t),0\!\!\leq \!\! t\!\leq \!\tilde{T}\})$ to (P2), it can be shown from (\ref{t_c}) and (\ref{P2c1}) that $\underset{0\leq t\leq \tilde{T}}{\max}\!t\!-\!t_{\mathrm{N}}(t)\!\leq\! \underset{i\in\{1,...,\tilde{N}+1\}}{\max}\!\tilde{t}_{i}^{\mathrm{I}}\!-\!\tilde{t}_{i-1}^{\mathrm{O}}\!\leq\! \bar{O}_{\mathrm{T}}$ holds, thus $(\tilde{T},\{\tilde{\mv{u}}(t),0\leq t\leq \tilde{T}\})$ is a feasible solution to (P1) and achieves the same objective value as (P2) with the solution $(\tilde{\mv{I}},\{\tilde{t}_i^{\mathrm{I}},\tilde{t}_i^{\mathrm{O}}\}_{i=1}^{\tilde{N}},  \tilde{T},\{\tilde{\mv{u}}(t),0\!\leq \!t\!\leq \!\tilde{T}\})$. Hence, the optimal value of (P1) is no larger than that of (P2). Therefore, (P1) and (P2) have the same optimal value, which completes the proof of Proposition \ref{prop_P1eq}.

\subsection{Proof of Proposition \ref{prop_connected}}\label{proof_prop_connected}
We prove Proposition \ref{prop_connected} by showing that for any feasible solution to (P2) denoted as $(\tilde{\mv{I}},\{\tilde{t}_i^{\mathrm{I}},\tilde{t}_i^{\mathrm{O}}\}_{i=1}^{\tilde{N}},\tilde{T},\{\tilde{\mv{u}}(t),0\leq t\leq \tilde{T}\})$ that does not satisfy the conditions in (\ref{tiI})--(\ref{T}), we can always find an alternative feasible solution to (P2) denoted as $(\tilde{\mv{I}},\{{t}_i^{\mathrm{I}},{t}_i^{\mathrm{O}}\}_{i=1}^{\tilde{N}},T,\{{\mv{u}}(t),0\!\leq\! t\!\leq \!{T}\})$ that satisfies the conditions in (\ref{tiI})--(\ref{T}) and achieves a smaller objective value of (P2) compared to the solution $(\tilde{\mv{I}},\{\tilde{t}_i^{\mathrm{I}},\tilde{t}_i^{\mathrm{O}}\}_{i=1}^{\tilde{N}},\tilde{T},\{\tilde{\mv{u}}(t),0\!\leq \!t\!\leq \!\tilde{T}\})$. Specifically, we first construct the same set of waypoint locations in ${\mv{u}}(t)$ as those in $\tilde{\mv{u}}(t)$, i.e., ${\mv{u}}_i^{\mathrm{I}}\!=\!\tilde{\mv{u}}(\tilde{t}_i^{\mathrm{I}}),{\mv{u}}_i^{\mathrm{O}}\!=\!\tilde{\mv{u}}(\tilde{t}_i^{\mathrm{O}}),i=1,...,\tilde{N}$. Then, we construct $\{{t}_i^{\mathrm{I}},{t}_i^{\mathrm{O}}\}_{i=1}^{\tilde{N}}$, $T$ and $\{{\mv{u}}(t),0\leq t\leq {T}\}$ according to (\ref{tiI})--(\ref{T}) based on the constructed $\{{\mv{u}}_i^{\mathrm{I}},{\mv{u}}_i^{\mathrm{O}}\}_{i=1}^{\tilde{N}}$. Note that since $(\tilde{\mv{I}},\{\tilde{t}_i^{\mathrm{I}},\tilde{t}_i^{\mathrm{O}}\}_{i=1}^{\tilde{N}},\tilde{T},\{\tilde{\mv{u}}(t),0\leq t\leq \tilde{T}\})$ satisfies the constraints in (\ref{P2c1}), it follows that ${\mv{u}}_i^{\mathrm{I}}\in \mathcal{C}_{I_i}$ and ${\mv{u}}_i^{\mathrm{O}}\in \mathcal{C}_{I_i}$ hold for all $i\in \{1,...,N\}$. Therefore, the constraints in (\ref{P2c1}) are also satisfied with the solution $(\tilde{\mv{I}},\{{t}_i^{\mathrm{I}},{t}_i^{\mathrm{O}}\}_{i=1}^{\tilde{N}},T,\{{\mv{u}}(t),0\leq t\leq {T}\})$ since $\mathcal{C}_{I_i}$ is a convex set, thus any point on the line segment between ${\mv{u}}_i^{\mathrm{I}}$ and ${\mv{u}}_i^{\mathrm{O}}$ also lies in $\mathcal{C}_{I_i}$. Moreover, it can be shown that $\underset{i\in \{1,...,\tilde{N}+1\}}{\max}\!t_{i}^{\mathrm{I}}\!-\!t_{i-1}^{\mathrm{O}}\!\leq\! \underset{i\in \{1,...,\tilde{N}+1\}}{\max}\!\tilde{t}_{i}^{\mathrm{I}}\!-\!\tilde{t}_{i-1}^{\mathrm{O}}\!\leq\! \bar{O}_{\mathrm{T}}$ holds, since the minimum time duration for the UAV to fly between two points with horizontal locations ${\mv{u}}_{i-1}^{\mathrm{O}}$ and ${\mv{u}}_i^{\mathrm{I}}$ is achieved by letting the UAV fly in a straight path as shown in (\ref{tiI})--(\ref{T}). This thus indicates that $(\tilde{\mv{I}},\{{t}_i^{\mathrm{I}},{t}_i^{\mathrm{O}}\}_{i=1}^{\tilde{N}},T,\{{\mv{u}}(t),0\leq t\leq {T}\})$ is a feasible solution to (P2). Furthermore, note that $T=\sum_{i=1}^{\tilde{N}+1} (t_i^{\mathrm{I}}-t_{i-1}^{\mathrm{O}})+\sum_{i=1}^{\tilde{N}} (t_i^{\mathrm{O}}-t_i^{\mathrm{I}})< \sum_{i=1}^{\tilde{N}+1} (\tilde{t}_i^{\mathrm{I}}-\tilde{t}_{i-1}^{\mathrm{O}})+\sum_{i=1}^{\tilde{N}} (\tilde{t}_i^{\mathrm{O}}-\tilde{t}_i^{\mathrm{I}})=\tilde{T}$ holds, since the minimum time duration for the UAV to fly between two points with horizontal locations ${\mv{u}}_{i-1}^{\mathrm{O}}$ and ${\mv{u}}_i^{\mathrm{I}}$, or ${\mv{u}}_{i}^{\mathrm{I}}$ and ${\mv{u}}_i^{\mathrm{O}}$, is achieved by letting the UAV fly in a straight path as shown in (\ref{tiI})--(\ref{T}), which are not satisfied by the solution $(\tilde{\mv{I}},\{\tilde{t}_i^{\mathrm{I}},\tilde{t}_i^{\mathrm{O}}\}_{i=1}^{\tilde{N}},\tilde{T},\{\tilde{\mv{u}}(t),0\leq t\leq \tilde{T}\})$. Therefore, $(\tilde{\mv{I}},\{{t}_i^{\mathrm{I}},{t}_i^{\mathrm{O}}\}_{i=1}^{\tilde{N}},T,\{{\mv{u}}(t),0\!\leq \! t\!\leq \!{T}\})$ achieves a smaller objective value of (P2) compared to $(\tilde{\mv{I}},\{\tilde{t}_i^{\mathrm{I}},\tilde{t}_i^{\mathrm{O}}\}_{i=1}^{\tilde{N}},\tilde{T},\{\tilde{\mv{u}}(t),0\!\leq \!t\!\leq \!\tilde{T}\})$, which thus completes the proof of Proposition \ref{prop_connected}.

\subsection{Proof of Lemma \ref{prop_waypoint}}\label{proof_prop_waypoint}
Note that with given $\mv{I}$, (P3-F) can be solved by solving $N+1$ parallel optimization problems, where each $i$th problem aims to minimize $\|{\mv{u}}_{i}^{\mathrm{I}}-{\mv{u}}_{i-1}^{\mathrm{O}}\|$ by optimizing ${\mv{u}}_i^{\mathrm{I}}$ and ${\mv{u}}_{i-1}^{\mathrm{O}}$ under the constraints in (\ref{P3c1})--(\ref{P3c4}). For any $i\in \{2,...,N\}$, it can be shown from the triangle inequality as well as (\ref{P3c3}) and (\ref{P3c4}) that $\|{\mv{g}}_{I_{i}}-{\mv{g}}_{I_{i-1}}\|=\|({\mv{u}}_{i}^{\mathrm{I}}-{\mv{u}}_{i-1}^{\mathrm{O}})+({\mv{u}}_{i-1}^{\mathrm{O}}-{\mv{g}}_{I_{i-1}})+({\mv{g}}_{I_{i}}-{\mv{u}}_{i}^{\mathrm{I}})\|\leq \|{\mv{u}}_{i}^{\mathrm{I}}-{\mv{u}}_{i-1}^{\mathrm{O}}\|+\|{\mv{u}}_{i-1}^{\mathrm{O}}-{\mv{g}}_{I_{i-1}}\|+\|{\mv{u}}_{i}^{\mathrm{I}}-{\mv{g}}_{I_{i}}\|\leq  \|{\mv{u}}_{i}^{\mathrm{I}}-{\mv{u}}_{i-1}^{\mathrm{O}}\|+2\bar{d}$ holds, which implies that $\|{\mv{u}}_{i}^{\mathrm{I}}-{\mv{u}}_{i-1}^{\mathrm{O}}\|\geq \max\{\|{\mv{g}}_{I_{i}}-{\mv{g}}_{I_{i-1}}\|-2\bar{d},0\}$ holds due to the non-negativeness of norm functions. Similarly, it can be shown that $\|{\mv{u}}_0-{\mv{g}}_{I_1}\|=\|({\mv{u}}_0^{\mathrm{O}}-{\mv{u}}_1^{\mathrm{I}})+({\mv{u}}_1^{\mathrm{I}}-{\mv{g}}_{I_1})\|\leq\|{\mv{u}}_1^{\mathrm{I}}-{\mv{u}}_0^{\mathrm{O}}\|+\bar{d}$ and $\|{\mv{u}}_F-{\mv{g}}_{I_N}\|=\|({\mv{u}}_{N+1}^{\mathrm{I}}-{\mv{u}}_N^{\mathrm{O}})+({\mv{u}}_N^{\mathrm{O}}-{\mv{g}}_{I_N})\|\leq\|{\mv{u}}_{N+1}^{\mathrm{I}}-{\mv{u}}_N^{\mathrm{O}}\|+\bar{d}$ hold, which implies that $\|{\mv{u}}_1^{\mathrm{I}}-{\mv{u}}_0^{\mathrm{O}}\|\geq \max\{\|{\mv{u}}_0-{\mv{g}}_{I_1}\|-\bar{d},0\}$ and $\|{\mv{u}}_{N+1}^{\mathrm{I}}-{\mv{u}}_N^{\mathrm{O}}\|\geq \max\{\|{\mv{u}}_F-{\mv{g}}_{I_N}\|-\bar{d},0\}$ hold. By further noting that the solution in (\ref{P3Fu1})--(\ref{P3Fu4}) yields $\|{\mv{u}}_{i}^{\mathrm{I}}-{\mv{u}}_{i-1}^{\mathrm{O}}\|=\max\{\|{\mv{g}}_{I_{i}}-{\mv{g}}_{I_{i-1}}\|-2\bar{d},0\},\ \forall i\in \{2,...,N\}$, $\|{\mv{u}}_1^{\mathrm{I}}-{\mv{u}}_0^{\mathrm{O}}\|= \max\{\|{\mv{u}}_0-{\mv{g}}_{I_1}\|-\bar{d},0\}$, and $\|{\mv{u}}_{N+1}^{\mathrm{I}}-{\mv{u}}_N^{\mathrm{O}}\|= \max\{\|{\mv{u}}_F-{\mv{g}}_{I_N}\|-\bar{d},0\}$, the optimal value of (P3-F) with given $\mv{I}$ is given in (\ref{O_TI}). This thus completes the {\hbox{proof of Lemma \ref{prop_waypoint}.}}

\vspace{-1mm}
\begin{spacing}{0.87}
\bibliographystyle{IEEEtran}
\bibliography{ICC2019_arXiv}
\end{spacing}
\end{document}